# Self-Driven Atomic Dispersion in Graphitic Layers

Zhaoxi Chen[a], Yulu He[b], Zhuoran Yao[b], Jian Liu[b], Jun Cai[b], Ziyi Fan[c], Wenjun Zhang[c], Lei Lei[a], Zupeng Chen[c*], Bo Yang[b], Zhi Liu[a*], Zhu-Jun Wang[b*]

[a] Center for Transformative Science (CTS), ShanghaiTech University, No. 393 Huaxia Rd, Shanghai 201210, China

[b] School of Physical Science and Technology (SPST) & Shanghai Key Laboratory of High-resolution Electron Microscopy, ShanghaiTech University, No. 393 Huaxia Rd, Shanghai 201210, China

[c] College of Chemical Engineering, Nanjing Forestry University, No.159 Longpan Road, Nanjing 210037, China

## Abstract

Carbon-supported single-atom catalysts maximize metal utilization, but how metal nanoparticles transform into isolated atoms within carbon remains unclear. We show that metal nanoparticles can undergo a self-driven dispersion process under hydrocarbon oxidation conditions, transforming into single atoms that are confined in carbon matrix. Using Pt-catalysed hydrocarbon oxidation as a model, we combine *operando* electron microscopy, near-ambient-pressure X-ray photoelectron spectroscopy and mass spectrometry to track coupled structural and chemical evolution. Graphitic carbon grows at step edges of Pt nanoparticle, continuously reconstructing Pt surface and generating undercoordinated sites for atom release. *In-situ* generated CO accumulates at the metal-carbon interface, weakening bonding and facilitating self-amplified atom release and migration. Defective carbon overlayers then trap, stabilize and transport liberated atoms, while oxidative etching preserves interfacial access of reaction-gas. Similar behaviour across other metals suggests a general atomization pathway for single-atom catalyst synthesis, yielding products with electrocatalytic hydrogen production activity beyond standard commercial benchmarks.

[*] Author to whom correspondence should be addressed: czp@njfu.edu.cn; liuzhi@shanghaitech.edu.cn; wangzhj3@shanghaitech.edu.cn.

## Introduction

Single-atom catalysts (SACs) on carbon supports have emerged as a compelling platform in heterogeneous catalysis because they maximize metal utilization while offering distinctive electronic structures and catalytic behaviours[1-9]. Despite rapid progress in the field, an important question remains open: how atomically dispersed metal sites emerge and stabilize within carbon matrices under realistic reaction conditions. Bottom-up synthetic strategies, including precursor anchoring, ligand removal and support-assisted trapping, have greatly expanded access to carbon-supported SACs[2,4,10]. However, the dynamic pathways through which isolated metal atoms form, migrate and persist under working conditions remain incompletely understood. Clarifying these processes would not only advance the rational synthesis of SACs, but also deepen our understanding of catalyst restructuring in reactive environments.

Top-down redispersion of metal nanoparticles (NPs) under reactive atmospheres has recently emerged as a promising pathway to atomically dispersed catalysts[8,11-16]. In such systems, gas-phase adsorbates can reversibly perturb metal cohesion, restructure surfaces and generate low-coordination sites, thereby enabling fragmentation into smaller clusters or even isolated atoms[15,17]. Yet the atomistic basis of this process remains poorly understood, especially for carbon-rich environments in which metal surfaces, carbon overlayers and reactive molecules are strongly coupled[18,19]. Carbon deposition is usually regarded as a deactivation pathway because it encapsulates active metal surfaces and blocks reactant access[20,21]. Whether this same carbon overlayer can instead participate in atomization, and if so through what mechanism, has remained largely unexplored. Real-time, spatially resolved evidence linking interfacial chemistry to NPs fragmentation and single-atom formation has been particularly scarce[1].

Here we show that carbon encapsulation can function not only as a deactivation process, but also as an active driver of metal atomization. We use platinum (Pt) NPs as a model system given Pt's technological significance as a vital catalyst for hydrocarbon oxidation requiring both high activity and thermal stability, such as catalytic combustion and automotive exhaust purification[8,22-25]. By integrating *operando* electron microscopy, near-ambient-pressure X-ray photoelectron spectroscopy (NAP-XPS) and online mass spectrometry (MS), together with first-principles calculations and molecular dynamics (MD) simulations, we track the coupled structural and chemical evolution of Pt in real time. We find that graphitic layers (GL) growth at Pt step edges progressively reconstructs the NP surface[26], generating undercoordinated sites susceptible to atom release. At the same time, carbon monoxide formed during hydrocarbon oxidation accumulates at the confined Pt-C interface, weakens interfacial bonding and lowers the barrier for metal detachment and migration[17,27]. The resulting GL do not merely encapsulate the metal; instead, through their defects and continuous restructuring, they capture, stabilize and transport the released Pt atoms.

This process establishes a self-amplifying feedback loop[28]. Newly dispersed Pt species enhance catalytic turnover and promote further CO formation, which in turn accelerates interfacial restructuring and continued atom release. Meanwhile, the dynamic competition between GL growth and oxidative etching preserves access to the buried interface and generates a defect-rich carbon network that sustains outward atomic migration[29-31]. In this way, the evolving Pt-C-gas interface functions as an autonomous atomization engine, coupling catalyst restructuring directly to catalytic chemistry. These observations provide a direct mechanistic link between atomic-scale interfacial dynamics and macroscopic catalytic behaviour, and reveal how atomically dispersed species can emerge spontaneously from carbon-encapsulated metal NPs under *operando* conditions.

Beyond Pt, we further show that this self-driven atomization behaviour extends across multiple noble metals, including Pd, Rh and Ru, indicating that the mechanism is not material-specific but rooted in a broader interfacial principle. The resulting carbon-confined SACs can also be synthesized at larger scale and exhibit strong electrocatalytic hydrogen-evolution performance, with the Pt-based material surpassing commercial Pt/C in both activity and durability[32]. More broadly, this work reframes carbon encapsulation from a deactivation phenomenon into a dynamic pathway for metal atomization, and establishes a general mechanistic framework for understanding and exploiting the emergence of carbon-supported SACs.

## Results and Discussion

To elucidate how metal NPs evolve under hydrocarbon oxidation conditions, we examined the coupled interactions between Pt and the reactive gas environment during alkane and alkene oxidation. These reactions drive structural evolution across an unusually broad range of length scales, from Ångström-scale surface reconstruction to mesoscale metal-$sp^2$-carbon interactions. By integrating *operando* transmission electron microscopy (TEM), NAP-XPS, environmental scanning electron microscopy (ESEM) and online MS, we directly tracked Pt NPs fragmentation together with porous GL growth and the emergence of dispersed Pt species. These observations reveal how catalyst restructuring, interfacial carbon formation and catalytic chemistry evolve together during atomization, and provide the mechanistic basis for scalable synthesis of carbon-confined SACs and evaluation of their electrocatalytic performance.

## Fragmentation and Self-Driven Atomic Dispersion

Coking is commonly expected to deactivate Pt NPs by encapsulating the metal surface and blocking access to active sites. Unexpectedly, under hydrocarbon oxidation conditions, bare Pt NPs instead progressively fragment and generate atomically dispersed Pt species within the growing graphitic shell. As illustrated in **Figure 1**a, this overall transformation proceeds through two coupled stages: particle fragmentation followed by atomic dispersion-ultimately generating a hierarchical $Pt^1$/Gr-$Pt^N$ architecture. We designate this novel hierarchical structure as $Pt^1$/Gr-$Pt^N$, where the residual metallic core ($Pt^N$) is enveloped by a graphitic shell (Gr) hosting atomically dispersed Pt ($Pt^1$).

To capture this transformation directly, we exposed annealed Pt NPs to an ethylene and oxygen mixture ($C_2H_4$:$O_2$ = 1:3) under 4000 Pa at 873 K within an environmental TEM cell. *Operando* scanning transmission electron microscopy (STEM) reveals that Pt NPs underwent fragmentation followed by atomic dispersion (**Figure 1**b). During this process, the reshaping NPs surfaces are encapsulated by GL (**Figure 1**c), consistent with previous observations of metal-catalyzed carbon growth[33-37]. The key observation is therefore not carbon encapsulation alone, but carbon encapsulation accompanied by active fragmentation and metal release.

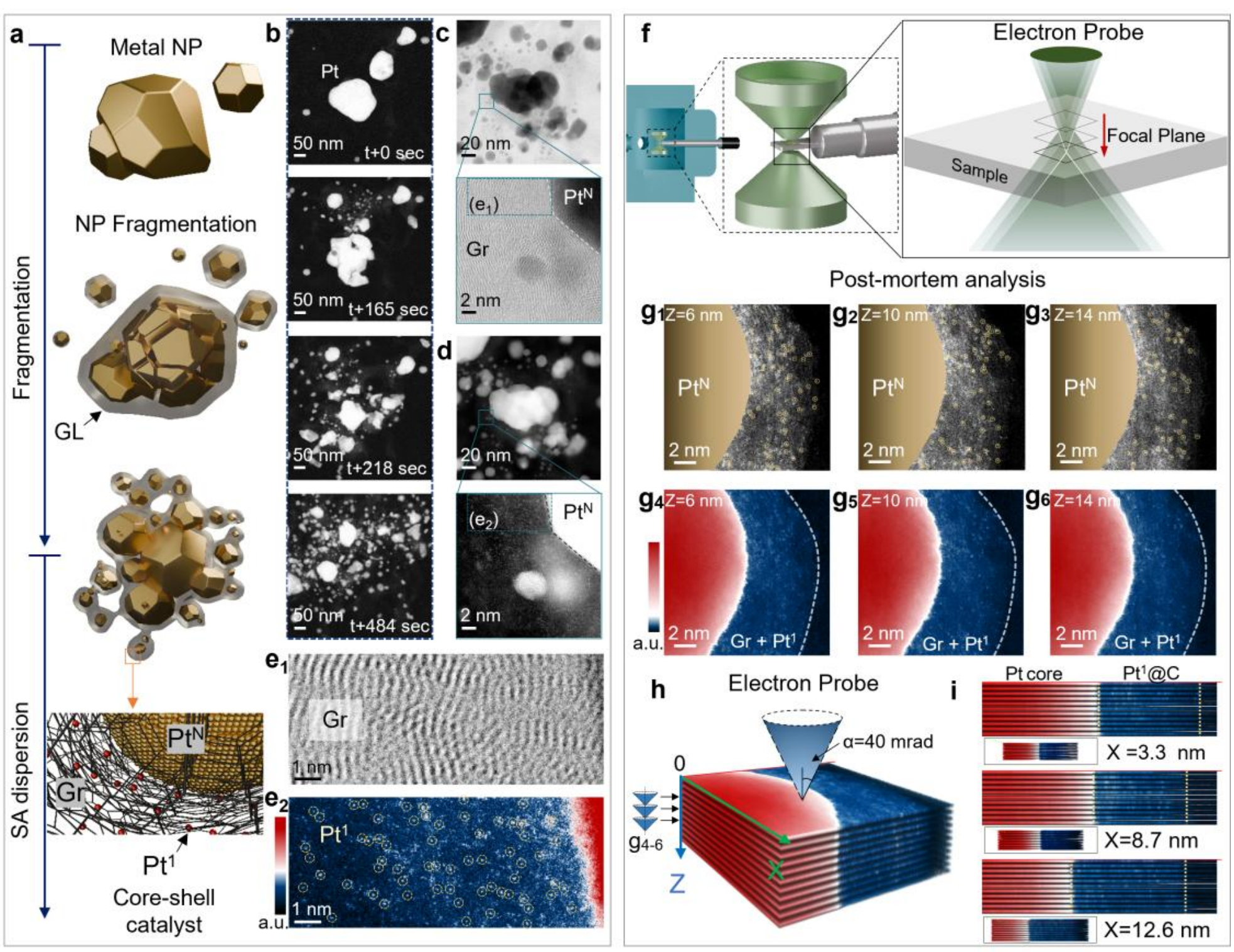

***Figure 1. Dynamic Transition of Pt NPs into Carbon-Supported SACs.*** *(a) Schematics of the fragmentation methodology for preparing the core-shell SACs through simultaneous GL growth and single atoms dispersion. (b) Operando STEM images showing the fragmentation of Pt metals under a $C_2H_4$ (1000 Pa) -$O_2$(3000 Pa) mixture at 873K. (c-e) STEM images displaying single atoms formation at the GL. Panels ($e_{1,\ 2}$) represent annular bright field (ABF) image and false-colored HAADF image in highlighting $Pt^1$ species marked in regions (c) and (d), respectively. (f) Post-mortem analysis of the single atoms position at the GL via a depth-sensitive sectioning method (See Methods). ($g_{1\text{-}6}$) Depth-sensitive STEM images captured at various focal planes showing the spatial distribution of single atoms within the GL. Panels $g_{1\text{-}3}$ and $g_{4\text{-}6}$ represent HAADF-STEM images, which have been subjected to contrast enhancement and false-coloring, respectively, to facilitate the clear visualization of Pt atoms located at the GL. (h, i) 3D-view of the stacked STEM slices at various depths. Images were aligned at the Pt-C interface.*

Atomic-resolution STEM further shows that this transformation culminates in the formation of isolated Pt species within the graphitic overlayer. High-angle annular dark field (HAADF)-STEM reveals the atomic feature of the resulting structures. Utilizing Z-contrast imaging, Pt atoms appear as distinct bright spots dispersed within the darker, lattice of the GL (**Figure 1**d, e). These carbon overlayers exhibit an interlayer spacing of ~0.35 nm[33-36]. To verify the three-dimensional location of Pt atoms, we employed depth-sensitive electron probe sectioning (**Figure 1**f). Remarkably, we observed singly dispersed Pt atoms within the GL. By scanning through the focal range, we pinpointed individual Pt atoms at varying depths within the GL, **Figure 1**f-h, meaning Pt atoms randomly distributed (Materials/Methods for imaging details). Electron diffraction and spectroscopic analysis (Energy Dispersive X-Ray Spectroscopy/Electron Energy Loss Spectroscopy, EDX/EELS) further confirm the GL structure and the presence of embedded, dispersed Pt atoms[38,39].

This observation upends the conventional view that carbon deposition leads solely to metal catalyst deactivation via site blocking[20,21,33,36]. Instead, we demonstrate that under tailored oxidative conditions, carbon encapsulation drives the extraction of single atoms from the core of Pt NPs. This fragmentation process establishes the foundation for self-driven atomic dispersion, the mechanistic details of which we elucidate in the following section.

**Multiscale *Operando* Investigation of Dynamic Fragmentation**

The self-driven atomic dispersion of Pt is closely coupled to hydrocarbon oxidation and GL growth, both of which continuously reshape the surface chemistry and morphology of the NPs. To follow these evolving states under reaction conditions, *operando* NAP-XPS was combined with online MS. With a probing depth of less than 1 nm[40], NAP-XPS provides direct access to the outermost chemical states of Pt and surface carbon species, while MS simultaneously tracks gas-phase reactants and products (**Figure 2**a-d). Together, these measurements enable real-time correlation of surface chemical evolution with catalytic turnover[41].

Upon heating to 673 K, adsorbed ethylene ($C_2H_4$*) desorbs from the Pt surface[42], marking the onset of catalytic ethylene oxidation (see **Figure 2**a C 1s spectra at 284.0 eV). This transition is evidenced by the appearance of $CO_2$ and $H_2O$ in the MS signal (**Figure 2**b) and corresponding gas-phase features in the O 1s spectra at 537.1 eV for $CO_2$ and 535.8 eV for $H_2O$. At the same time, the C 1s region shows a peak at 284.1 eV, assigned to hydrocarbon-derived C-H species formed during ethylene dissociation[43,44], along with a peak at 289.2 eV corresponding to carbonate species[45]. In the O 1s spectra, hydroxyl species react with dissociated hydrogen to form $H_2O$, which desorbs at elevated temperature. Between 673 K and 773 K, oxidation of carbonaceous intermediates continues, generating both CO and $CO_2$, while carbonate, oxygen-containing species and hydrocarbon residues remain detectable on the surface (see evolution of C 1s and O 1s spectra in **Figure 2**a). Over this temperature range, the Pt 4f signal progressively attenuates (see evolution of Pt 4f spectra in **Figure 2**a, d), consistent with increasing surface coverage by hydrocarbon-derived residues and carbonates that increasingly block the Pt surface.

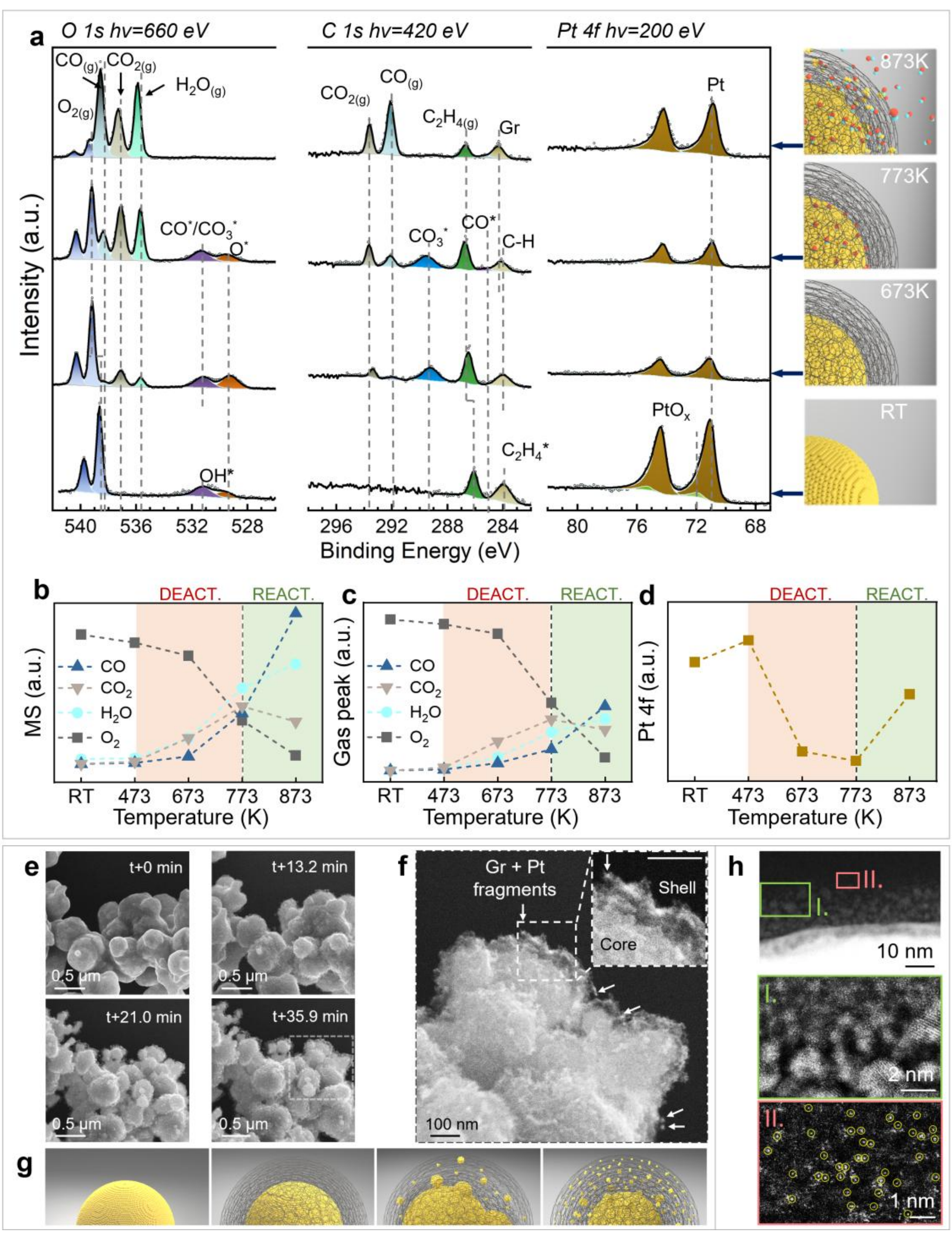

***Figure 2. Spectroscopic and Environmental SEM Investigations of Dynamic Catalyst Fragmentation.*** *(a) NAP-XPS measurements on Pt NPs under $C_2H_4$ (25 Pa) and $O_2$ (75 Pa). O 1s, C 1s, and Pt 4f spectra were collected from 298K to 873K with the photon energies of 660 eV, 420 eV, and 200 eV, respectively. The inserts show schematics of carbon*

*encapsulation and surface species. (b, c) The evolution of gas phase species as a function of temperature extracted from the MS (b) and XPS (c). (d) The Pt 4f peak intensity at different temperatures. (e) ESEM observations showing the Pt fragmentation behavior under ethylene oxidation at 873K. A 100% ethylene gas (100 Pa) was applied at t = 0 min. For t ⩽35.9 min, the gas composition gradually shifts to $C_2H_4$ (25 Pa) and $O_2$ (75 Pa). (f) An enlarged view of the region marked in (e). Scale bar at inset: 100 nm. (g) Schematics of Pt fragmentations in successive stages. (h) Post-mortem STEM characterizations of the ESEM specimen after ethene oxidation reaction.*

A distinct transition occurs at 873 K. As shown by MS (**Figure 2**b), CO becomes the dominant gas-phase product, while NAP-XPS detects a pronounced increase in gas-phase CO intensity, with corresponding signals at approximately 290.5 eV in C 1s and 536.2 eV in O 1s (**Figure 2**a, c). Concurrently, the C 1s peak shifts to 284.4 eV, consistent with the formation of GL[44]. Most notably, the Pt 4f signal exhibits a marked recovery after its earlier attenuation (**Figure 2**d), indicating renewed exposure of Pt species within the XPS probing depth. This recovery is consistent with the outward redistribution of Pt during fragmentation and atomic dispersion, in agreement with the *operando* STEM observations in **Figure 1**.

Taken together, these *operando* measurements identify 873 K as a critical threshold at which the catalyst transitions from progressive encapsulation and passivation to renewed activity accompanied by Pt redistribution. The simultaneous increase in CO production, GL formation and Pt re-exposure suggests the onset of a self-amplifying process in which newly generated dispersed Pt species contribute to further hydrocarbon activation and CO formation[46,47]. In this picture, CO is not merely a reaction product, but part of a coupled interfacial response associated with catalyst restructuring. The combined NAP-XPS and MS data therefore point to a thermal-activated transition in which surface passivation gives way to dynamic fragmentation and renewed catalytic turnover.

To connect these chemical observations with structural evolution over larger length scales, ESEM was used to follow the mesoscale response of Pt NPs under reactive conditions. ESEM imaging at 873 K reveals progressive encapsulation, particle reshaping and fragmentation during ethylene oxidation (**Figure 2**e, f), as summarized schematically in **Figure 2**g. Post-reaction atomic-resolution STEM further confirms that this mesoscale restructuring is accompanied by the formation of atomically dispersed Pt species within the graphitic shell (**Figure 2**h). Unlike the relatively uniform GL typically observed on extended Pt surfaces[34-36], the structures formed here are spatially heterogeneous, with graphitic carbon, residual clusters and dispersed Pt species coexisting, reflecting the non-equilibrium restructuring of Pt NPs under reaction conditions.

Together, these multiscale *operando* observations establish a direct connection between surface chemistry, gas-phase reaction kinetics and structural evolution. They identify 873 K as the point at which the reaction pathway changes qualitatively: carbon encapsulation is no longer associated solely with deactivation, but becomes coupled to fragmentation, Pt redistribution and renewed catalytic activity. This transition provides the experimental basis for the interfacial mechanism examined in the following section. The same behaviour is observed over a broad total-pressure range from $10^1$ to $10^4$ Pa, indicating that the self-driven dispersion process is robust across reaction environments relevant to both *operando* observation and scalable synthesis.

**CO-Intercalation, Interface Restructuring, and Self-Amplifying Fragmentation**

Under reactive conditions, the chemical potentials of product gases fundamentally alter the physicochemical properties of NPs surfaces[48]. In the present system, however, the graphitic overlayer creates a confined microenvironment at the metal-carbon interface that fundamentally changes this response[19,49]. Within such confined regions, CO can accumulate at undercoordinated Pt step-edge sites, where it weakens metal-metal cohesion and promotes surface disintegration[50-52]. Rather than preserving the original stepped morphology, this process drives the transformation of monoatomic steps into nanoscale protrusions and roughened interfacial facets[17,50], as observed in **Figure 3**e-i (highlighted by colored arrows). These observations suggest that interfacial confinement amplifies CO-mediated restructuring and identifies CO-induced reconstruction of the Pt-C interface as a key driver of self-driven atomic dispersion.

To test this directly, *operando* STEM was synchronized with online MS during ethylene oxidation in a $C_2H_4/O_2$ mixture (1:3, total pressure 4000 Pa). As shown in **Figure 3**a-c, the carbon-encapsulated Pt NP remains morphologically static at 773 K between $t_1$ and $t_2$, while the MS signal shows declining product intensities, consistent with catalyst deactivation caused by graphitic encapsulation (**Figure 3**c, "DEACT." zone). A clear transition occurs as the temperature is increased from 823 to 873 K. During this interval, *operando* STEM reveals the spontaneous disintegration of the parent NP into smaller clusters beginning at $t_3$, whereas MS simultaneously records reversal of the activity decline, a pronounced exothermic event (marked by blue ▼ and vertical yellow dashed line), and accelerated reactant consumption together with increased product formation (**Figure 3**c, "REACT." zone). The onset of fragmentation thus coincides with catalyst reactivation and with the emergence of strong CO production, linking interfacial restructuring directly to macroscopic catalytic behaviour.

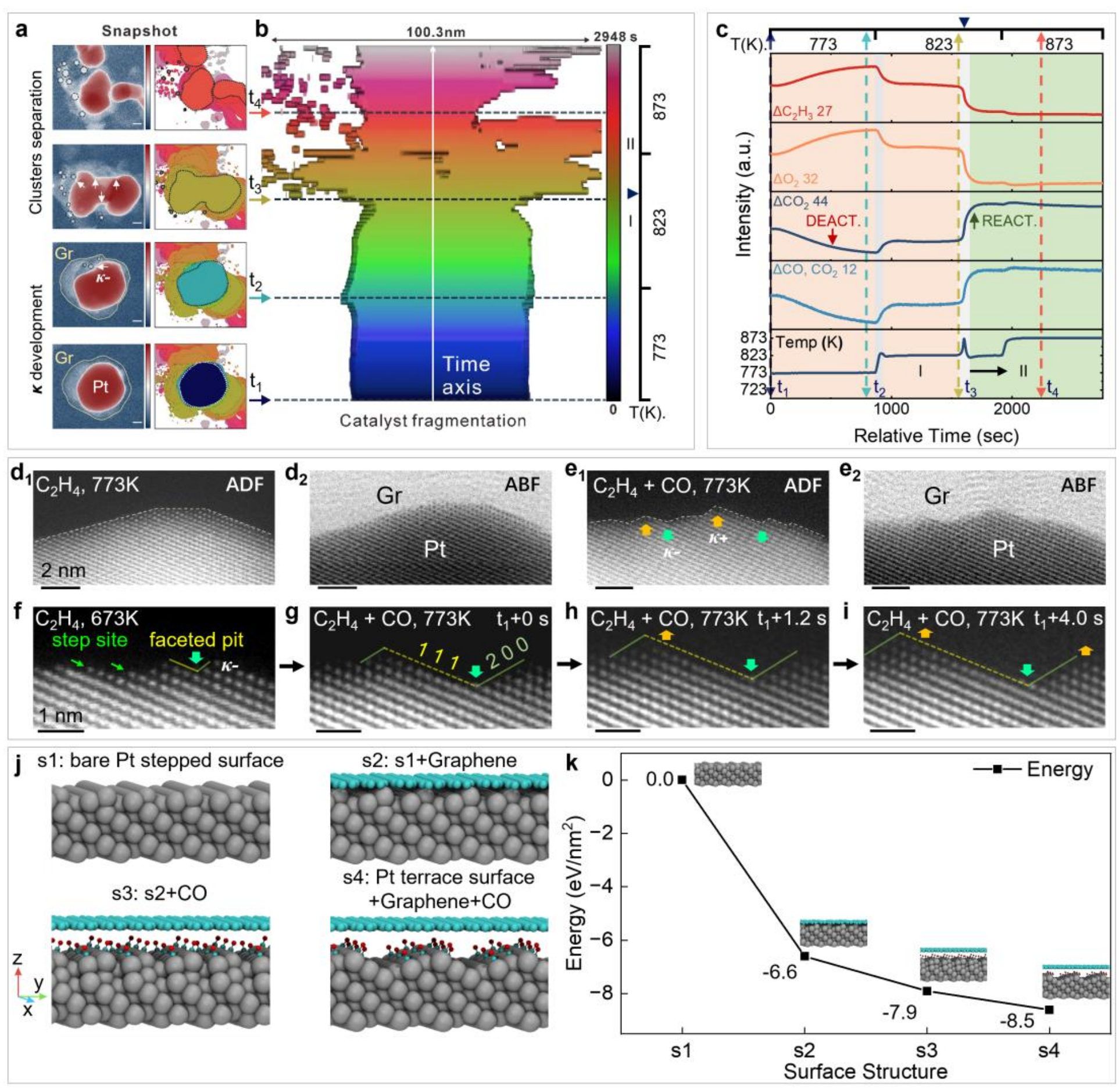

***Figure 3. Real-Space Observations of Dynamic Pt Fragmentation Behavior during Metal-Catalyzed GL Growth.*** *(a, b) Operando STEM observations reveal the dynamic disintegration pathway of Pt NPs into smaller clusters within an ethylene and oxygen mixture ($C_2H_4$:$O_2$ = 1:3) under 4000 Pa. Yellow dashed contours demarcate the carbon overlayer coverage on Pt surfaces, while white arrows identify concave surface features.*

*The time-stack in (b) illustrates the contours of segmented Pt NPs. Scale bar in (a): 10 nm. (c) Online MS data demonstrates enhanced ethylene oxidation reactivity following catalyst fragmentation. (d-i) Comparative STEM observations examine CO-induced fragmentations at the Pt-C interface. The morphology of NPs at the Pt-C interface is compared under ($d_{1,\ 2}$) $C_2H_4$ and ($e_{1,\ 2}$) ($C_2H_4$ + CO) mixture at 773K (see Methods). Arrows in (e) denote nano-protrusions with extended facets. (f-i) Atomically-resolved STEM images depict the reconstructions of mono-atomic steps and the formation of nano-protrusions at the Pt-C interface under the $C_2H_4$ + CO mixture. (j, k) Surface energy evolution during CO-induced reconstruction of stepped Pt(111)-(200) surfaces. The pristine Pt (111) surfaces with (200) type step facets serves as the reference state. Progressive energy minimization drives surface fragmentation into mixed (111)/(200) facets***.**

This transition points to a self-amplifying fragmentation process. Newly generated Pt species and small Pt clusters provide highly active sites for hydrocarbon activation, thereby increasing local CO formation. The elevated CO concentration, in turn, further promotes interfacial reconstruction and continued fragmentation of the parent particle. In this way, catalyst restructuring and catalytic turnover become dynamically coupled. Importantly, this CO-driven fragmentation remains robust over a broad range of $C_2H_4/O_2$ ratios and pressures, whereas high oxygen partial pressures accelerate more extensive NP disintegration. At the same time, the graphitic shell does not merely confine the metal: it also acts as a dynamic stabilizing matrix. Detached Pt fragments in the 2-6 nm range are captured within van der Waals "nano-pockets" formed by flexible GL, which suppresses re-aggregation. *Operando* STEM further shows that these confined fragments can themselves undergo outward dispersion and secondary fragmentation, rather than following the classical sintering pathway[53]. Meanwhile, oxidative etching of the graphitic shell[54] generates a porous structure that maintains gas access to the buried Pt-C interface, thereby sustaining continued restructuring.

The role of CO was further isolated by comparing Pt-C interfacial dynamics under pure ethylene and under mixed CO/$C_2H_4$ atmospheres. Under pure ethylene at 773 K, carbon encapsulation proceeds through hydrocarbon dehydrogenation at Pt step edges[34,55-57], which act as nucleation sites for graphitic growth (**Figure 3**d). In the absence of CO, the interface evolves towards strain-relieved morphologies dominated by close-packed Pt(111) terraces and stable step terminations beneath the graphitic shell[34,35]. By contrast, introducing CO together with $C_2H_4$ (CO:$C_2H_4$ = 2:1) dramatically alters the interfacial dynamics. As shown in **Figure 3**e-i, CO induces pronounced step mobility and the formation of ~2 nm protrusions directly beneath the graphitic overlayer. High-resolution time-lapse imaging (**Figure 3**f-i) reveals that these protrusions evolve into rough, mixed Pt(111)/(200) facets at the Pt-C interface. This behaviour is qualitatively different from the smooth, strain-accommodated morphology observed in the absence of CO[34,35], providing direct experimental evidence that CO destabilizes the encapsulated Pt surface and drives interfacial roughening (**Figure 3**e).

Density Functional Theory (DFT) calculations support this interpretation. As summarized in **Figure 3**j, k, progressive structural evolution from a stepped Pt(111)/(200) surface to a reconstructed interface is accompanied by sequential energy lowering. Graphene adsorption onto the stepped Pt surface stabilizes the interface, and subsequent CO intercalation lowers the energy further through Pt-CO chemisorption[17] together with weakening of Pt-C coupling[58]. Most importantly, reconstruction into interwoven Pt(111)/(200) terraces with alternating facet orientations yields an additional energetic reduction, indicating that the roughened, protruding morphology is thermodynamically favoured once CO is confined at the interface. The experimentally observed conversion of monoatomic steps into rough faceted protrusions is therefore consistent with a CO-stabilized interfacial reconstruction pathway.

Further insight is provided by DFT and molecular dynamics simulations of the confined Pt-C interface. Because graphene binds only weakly to Pt, thermal fluctuations permit local decoupling of the overlayer[59], enabling CO to intercalate into the confined interface. Electron-density-difference analysis shows that CO intercalation weakens Pt-C interactions, while molecular dynamics simulations show spontaneous structural relaxation of the interface upon CO insertion. Taken together with *operando* STEM, these calculations support a mechanism in which CO accumulation at the confined Pt-C interface weakens interfacial bonding, enhances step mobility and drives the formation of protrusions and detached fragments. Flexible GL then capture and stabilize these fragments, while continued CO generation and shell restructuring sustain the fragmentation process. This coupled interfacial response establishes the mechanistic basis for the self-amplifying conversion of Pt NPs into stable carbon-confined SACs.

**Defect-Mediated Atomic Dispersion: Sculpting Metal Clusters to Single Atoms**

Above results establish that GL serve as dynamic substrates for extracting $Pt^1$ from encapsulated NPs. *Operando* observations further reveal CO-induced surface fragmentation (see **Figure 3**g-i highlighted by green and yellow arrows) where two defect-mediated dispersion pathways emerge, both governed by interface-type defects. First, in-plane $Pt^1$ detachment proceeds via intercalation at graphite terminations. This process is facilitated by CO-induced surface reconstruction of Pt NPs, which generates low-coordination sites and nano-protrusions (**Figure 4**a, highlighted by white arrows). MD simulations in **Figure 4**b confirm that under identical Pt and carbon atomic configuration, the CO induced low-coordinate Pt sites (roughening of NPs surfaces via nano-protrusions formation) significantly enhances the kinetics of this mechanism (see **Figure 4**c). Second, out-of-plane $Pt^1$ dispersion occurs through defect channels within the GL stack. *Operando* imaging reveals that atomic-scale discontinuities including discontinued edges (**Figure 4**d), dislocations and stacking faults within the GL act as anchoring sites[60]. MD simulations of thermally annealed graphene multilayers with laterally disrupted edges (**Figure 4**e) reveal spontaneous structural reorganization. Through controlled heating-cooling cycles, edge carbon atoms reconfigure to eliminate high-energy defects, such as dangling bonds and strained rings. Discontinued GL undergo spontaneous self-mending, forming rigid, foam-like dislocation networks perpendicular to the Pt surface (**Figure 4**e). These cross-linked channels enable $Pt^1$ diffusion vertically through the GL stack (**Figure 4**f), with $Pt^1$-defect complexes exhibiting significantly enhanced stabilization energies (-5.33 eV vs. -1.17 eV on pristine graphene).

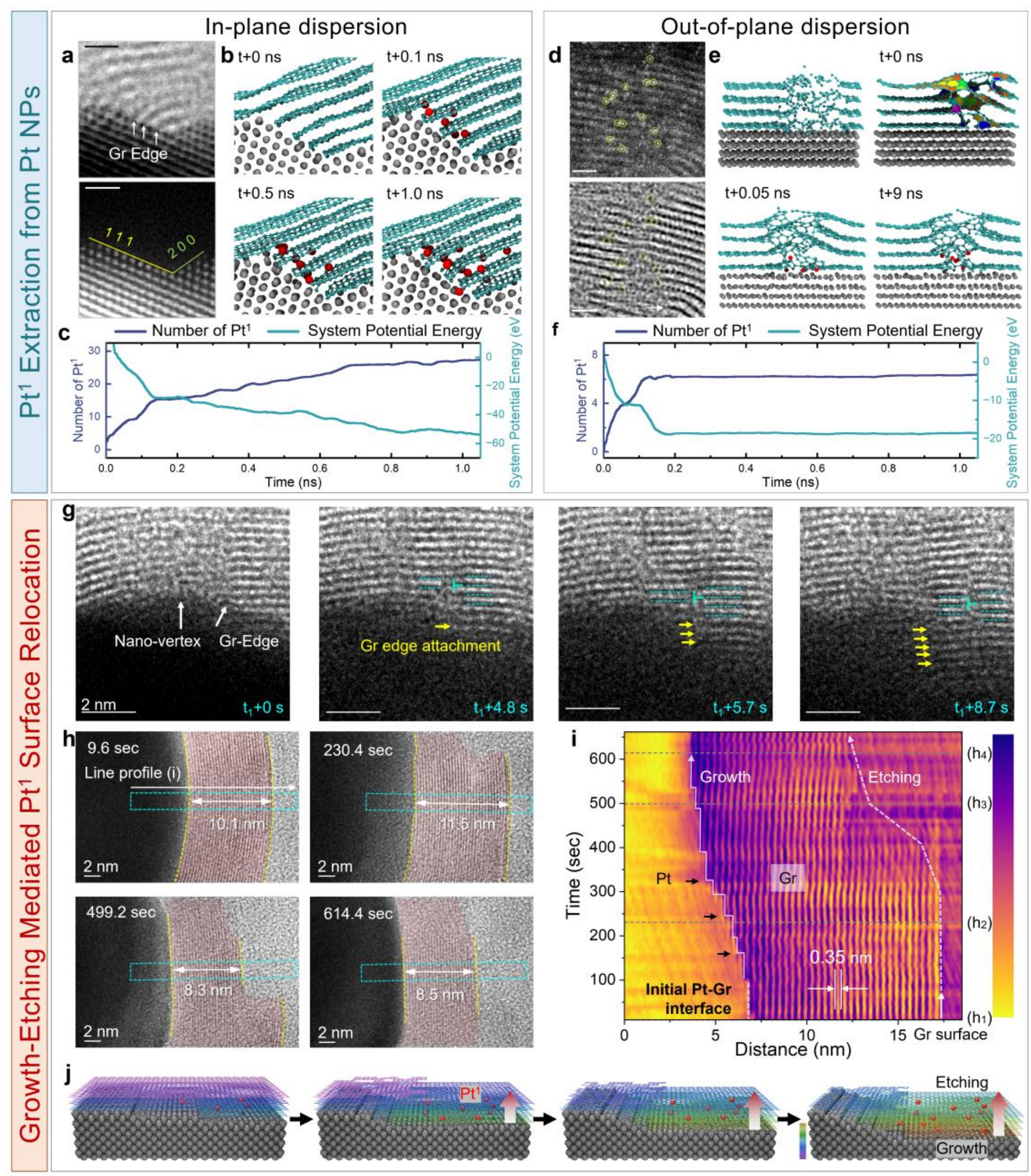

***Figure 4. Mechanisms of Defect-Mediated Single-Atom Dispersion during GL Growth.*** *(a) STEM images showing the GL terminations on the Pt(111) terrace. The arrow marks the lattice fringe of graphite (scale bar: 1 nm). (b) MD simulations demonstrating the dispersion of $Pt^1$ through the graphite edges. (d) STEM images showing the $Pt^1$ anchored by open channels in GL (scale bar: 1 nm). (e) MD simulations demonstrating the dispersion of $Pt^1$ to a disorderly cross-linked carbon defect. (c, f) The corresponding numbers of dispersed $Pt^1$ and change in system potentials within 1 ns in (b) and (e). (g) Operando TEM images showing the defects formation during GL growth on the Pt surface. The nucleation of Gr segments at the proximity of a Pt nano-protrusion under a $C_2H_4$-$O_2$ (1:3) mixture is highlighted by yellow arrows. The blue markers show the formation of discontinuous carbon edges during growth. (h, i) Simultaneous GL growth and etching on Pt NPs. (h) Time series of operando TEM images demonstrating the Pt surface step-mediated GL growth by segregation. The evolution of TEM line profiles transverse from the Pt-C interface to the carbon surface is depicted in (i) (blue box refers to the line profile region with a width of 2 nm; color mask and lines are used to guide the eye). (j) Schematics of the conveyor-like process for $Pt^1$ (red spheres) surface relocation through GL growth-etching regeneration. Red spheres represent single $Pt^1$, while grey spheres denote the $Pt^N$. The red gradient arrows indicate the transportation process of $Pt^1$ from the interface onto the GL. Different colors*

*represent different layers of Gr.*

The formation of these defects is spatially coupled to localized strain fields generated by Pt NP surface protrusions, which induce inelastic deformation within the overlaying GL[30]. Resultant defects facilitate reactant gas permeation while simultaneously initiating GL growth at the Pt-C interface proximal to protrusion sites (**Figure 4**g, h). A dynamic equilibrium between GL growth and etching (**Figure 4**i, j) drives continuous mass transport: nascent carbon layers formed at the Pt-C interface extrude and entrain $Pt^1$ species toward the outermost GL surface. This conveyor-like process operates synergistically with the aforementioned dispersion pathways, progressively relocating $Pt^1$ from the NP core to catalytically active surface positions of $Pt^1$/Gr-$Pt^N$.

The synergistic interplay between CO intercalation and GL growth drives continuous interface restructuring and defect proliferation in the evolving overlayers[48]. Unlike the stagnant encapsulation observed in pure alkane systems[35], ethylene oxidation over Pt NPs drives continuous surface reconstruction and shape evolution, creating a self-sustaining $Pt^1$ dispersion pathway. During this progressive self-driven dispersion, internal carbon segregation competes with external GL oxidation/etching[61,62]. Complementary Raman spectroscopy of defect-rich graphite quantifies structural evolution.

Theoretical simulation confirms the critical role of CO in $Pt^1$ dispersion. Previous investigations have shown that CO can readily intercalate the Pt-C interface, enhancing the adsorption of CO and oxygen species[63]. The adsorbates on Pt NPs weaken the metal catalyst coupling with GL and fragment the Pt NPs surface into tiny clusters[17,64]. DFT calculations further quantify stabilization energies for $Pt^1$-CO complexes on pristine graphene surface. The energy difference as compared to unbound Pt atoms is lower than the adatom formation energy on Pt terraces, steps, and kinks (1.2-2.6 eV)[50]. These energetically favored configurations, coupled with the dynamic migration of defects during GL growth, establish thermodynamic drivers for stabilizing interfacial $Pt^1$ at GL defect sites.

The mechanistic framework for SACs formation via CO mediated metal-carbon interactions was further validated across multiple catalytic systems: Rh- and Pd-catalyzed ethylene oxidation, alongside Pt-catalyzed propane oxidation. Critically, all systems exhibited identical fragmentation signatures: including nano-protrusion formation at metal-carbon interfaces and progressive atomic dispersion to defect sites. This consistent behavior demonstrates the inherent universality of the bond-cleavage mechanism, where strong metal-substrate interactions overcome intrinsic metal-metal cohesion energies to enable atomic separation.

The phenomenon manifests specifically during alkane/alkene oxidation cycles where two conditions converge: i) GL formation at temperatures $\geqslant$773 K and ii) enrichment of CO gas at the metal-carbon interface. Notably, Rh and Pd catalysts displayed fragmentation kinetics comparable to Pt during ethylene oxidation at low oxygen partial pressures, with *operando* imaging confirming simultaneous GL growth and metal disintegration. This ubiquitous mechanism redefines carbon encapsulation from a deactivation pathway to a dynamic atomization engine - where the GL conventionally blamed for catalyst deactivation instead become active participants in generating stabilized SACs.

**Scalable Synthesis and Electrocatalytic Performance**

To translate the atomic-scale dispersion mechanism into practical materials synthesis, an integrated reaction platform was developed for the scalable production of $X^1$/Gr-$X^N$ catalysts ($X$ = Pt, Pd, Rh, Ru). This system bridges *operando* TEM observations with macroscopic synthesis by reproducing, in a dedicated reaction chamber, the reactive nano-environment identified in the TEM nanoreactor (**Figure 5**a-c). In this way, the atomic-dispersion conditions established at the nanoscale can be directly transferred to gram-scale catalyst preparation.

As a proof of concept, the previously identified dispersion conditions were applied to synthesize $Pt^1$/Gr-$Pt^N$. SEM reveals a clear morphological transformation after reaction: compared with the parent Pt NPs (**Figure 5**d), the as-synthesized $Pt^1$/Gr-$Pt^N$ exhibits uniformly distributed GL with dispersed Pt clusters at the outmost surface (**Figure 5**e, f). High-resolution STEM further confirms the formation of the targeted structure by directly resolving isolated Pt atoms within the graphitic shell (**Figure 5**g).

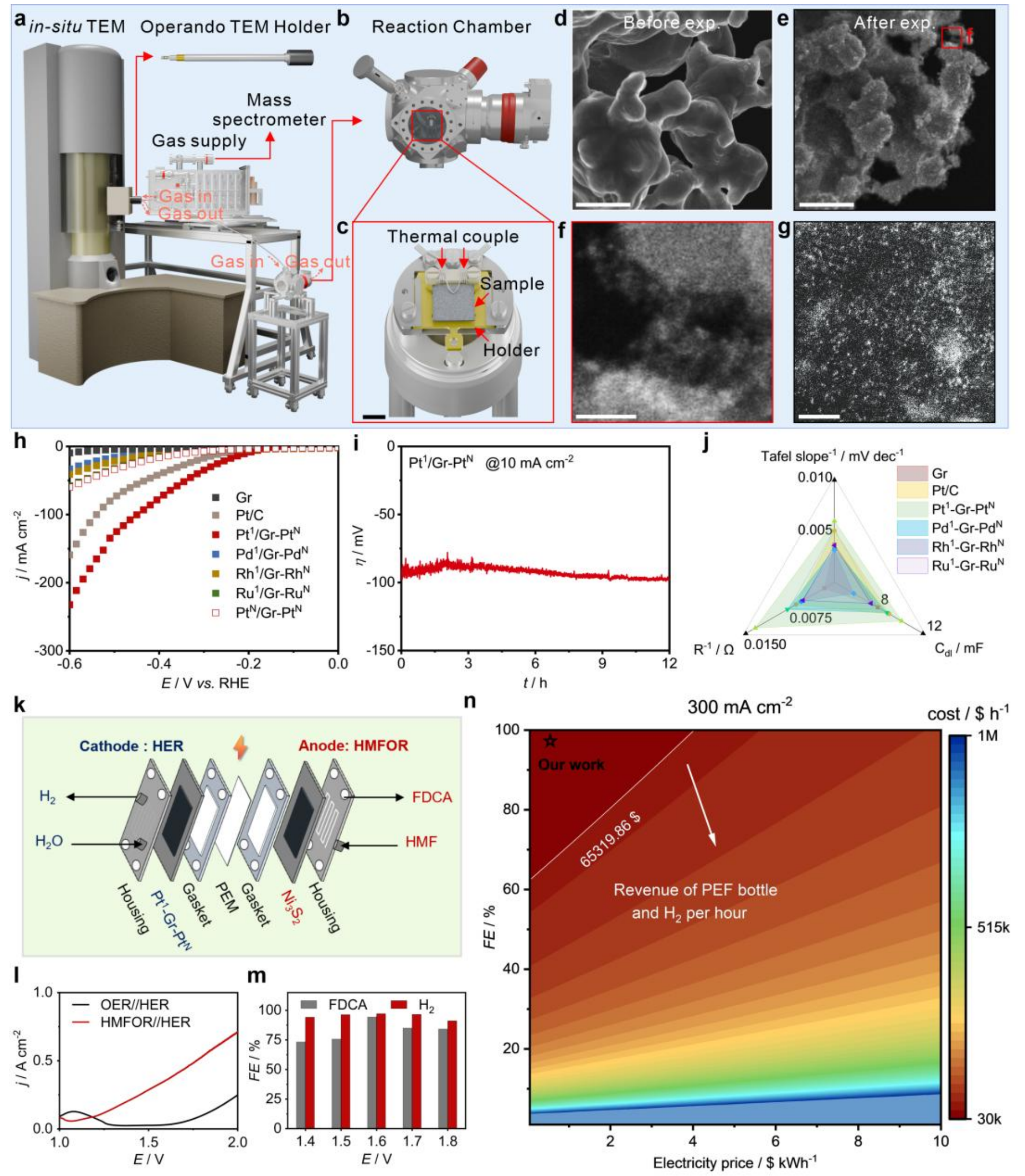

***Figure 5.*** **Scalability and Performance Evaluation of Self-driven Dispersed SACs.** *(a) The integrated reaction system for scalable production of SACs combining the operando TEM stage, a shared gas-supply module, and the macro-reactor. The gas-supply module delivers identical reactive atmospheres (alkane/alkene and oxygen mixtures) simultaneously to both the TEM holder and the macro-reaction chamber. (b, c) The reaction chamber and sample holder. Within the chamber, a customized sample holder provides real-time temperature feedback. Gas pressure and flow rates are independently regulated using a controlled leak valve and Mass Flow Controllers, respectively. The pentagram symbol marks the position of the sample. (d, e) SEM micrographs of Pt NPs comparison of sample morphologies before in (d) and after SACs production in (e). (f) High-resolution SEM images characterization of the local sample in (e). (g) HAADF images revealing the $Pt^1$ distributed within the GL. Scale bars:5 mm (c), 1 μm (d, e), 20 nm (f), 2 nm (g). (h) HER polarization curves of Gr, Pt/C, and $X^1$/Gr-$X^N$ (where "X" denotes Pt, Pd, Rh and Ru). (i) Time-dependent overpotential curve of $Pt^1$/Gr-$Pt^N$ at 10 mA $cm^{-2}$. (j) Tafel slopes, $C_{dls}$, and EIS-Nyquist plots of Gr, Pt/C, and $X^1$/Gr-$X^N$. (k) Schematic illustration of the HMFOR//HER MEA flow reactor. (l) LSV curves of $Pt^1$/Gr-$Pt^N$ (HER) and $Ni_2S_3$ (HMFOR) in MEA system. (m) Voltage-dependent performance of MEA system. (n) TEA of plant-gate levelized cost of producing PEF bottle and $H_2$ per hour.*

The electrocatalytic hydrogen-evolution reaction (HER) performance of $X^1$/Gr-$X^N$ ($X$ = Pt, Pd, Rh, Ru) was then evaluated. Linear sweep voltammetry (LSV) shows that all $X^1$/Gr-$X^N$ catalysts deliver higher current densities than reduced graphene oxide (Gr). Among them, $Pt^1$/Gr-$Pt^N$ exhibits the best HER performance, surpassing commercial Pt/C with an overpotential of 206 mV at 10 mA $cm^{-2}$, compared with 260.6 mV for Pt/C and 606.2 mV for Gr (**Figure 5**h). Specially, $Pt^1$/Gr-$Pt^N$ shows excellent durability, maintaining stable operation throughout a 12 h chronopotentiometry test at 10 mA $cm^{-2}$, moreover, the overpotential is much lower than that of other control samples (**Figure 5**I). To probe the origin of this enhancement, further kinetic and interfacial analyses were performed. $Pt^1$/Gr-$Pt^N$ shows a lower Tafel slope (171.4 mV $dec^{-1}$) than either Gr (278 mV $dec^{-1}$) or Pt/C (204.1 mV $dec^{-1}$), indicating accelerated HER kinetics due to atomic-scale dispersion Pt (**Figure 5**j). As a reference, Pt NPs confined within GL but without extensive atomic dispersion ($Pt^N$/Gr-$Pt^N$) were also prepared and tested. Their HER activity is slightly lower than that of $Pt^1$/Gr-$Pt^N$ and Pt/C, but remains substantially higher than that of Gr, with Tafel analysis showing a similar trend.

Double-layer capacitance measurements further show that $Pt^1$/Gr-$Pt^N$ possesses a higher $C_{dl}$ (9.9 mF $cm^{-2}$) than both Gr (7.8 mF $cm^{-2}$) and Pt/C (8.8 mF $cm^{-2}$), consistent with a larger electrochemically accessible surface population (**Figure 5**j). Electrochemical impedance spectroscopy (EIS) reveals the lowest charge-transfer resistance for $Pt^1$/Gr-$Pt^N$ (76.9 Ω), markedly below that of Gr (601.6 Ω) and Pt/C (158.3 Ω), confirming faster interfacial charge transfer (**Figure 5**j). By contrast, $Pd^1$/Gr-$Pd^N$, $Rh^1$/Gr-$Rh^N$, and $Ru^1$/Gr-$Ru^N$ exhibit lower $C_{dls}$ and larger Tafel slopes than Gr, but substantially reduced charge-transfer resistances. This suggests that, for these catalysts, the dominant contribution lies in improved electrical conductivity of the graphitic framework and more efficient interfacial charge transfer, rather than in a higher density of active sites or intrinsically faster HER kinetics.

To further evaluate the practical potential of the catalyst, the oxygen-evolution reaction (ORR) was replaced with the oxidation of biomass-derived 5-hydroxymethylfurfural (HMFOR), thereby coupling hydrogen production to the generation of the value-added product 2,5-furandicarboxylic acid (FDCA)[65,66]. In a membrane-electrode-assembly (MEA) flow reactor integrating HER and HMFOR (**Figure 5**k), the system reached an industrially relevant current density of 300 mA $cm^{-2}$ at only 1.52 V, representing a 530-mV reduction relative to conventional water splitting (**Figure 5**l). The activity shows a volcano-type dependence on voltage, and at 1.6 V the Faradaic efficiencies (FE) reach 94.3% for FDCA and 97.1% for $H_2$, corresponding to a combined electron utilization of 191.4% (**Figure 5**m). Techno-economic analysis (TEA) further indicates that profitability at 300 mA $cm^{-2}$ requires a FE above 62% and an electricity cost below \$4 $kWh^{-1}$. Under an industrial electricity price of \$0.08 $kWh^{-1}$, the system is projected to deliver a net profit of \$20,613 (**Figure 5**n). Together, these results demonstrate that the self-driven dispersion strategy is not only scalable, but also capable of generating high-performance catalysts for practical hydrogen production and coupled biomass upgrading.

## Conclusion

This work establishes a general top-down route for the scalable synthesis of carbon-supported single-atom catalysts by exploiting the dynamic evolution of the metal-gas-carbon interface. Rather than acting solely as a deactivation pathway, coking is shown here to drive the spontaneous conversion of metal nanoparticles into stable, carbon-confined single atoms under oxidative hydrocarbon conditions. This transformation arises from a confinement-enabled, self-amplifying process in which graphitic-layer growth and interfacial CO accumulation cooperatively promote surface disintegration, atomic detachment and migration into defect-rich carbon environments. Supported by *operando* measurements and theoretical calculations, this mechanism links catalyst restructuring directly to catalytic chemistry. Beyond Pt, this self-driven dispersion phenomenon is generalizable to a broader range of noble metals (Pd, Rh, Ir and Ru) and other transition metals (Re, Ni and Co). Furthermore, the resulting materials deliver robust electrocatalytic hydrogen-evolution performance, with the Pt-based catalysts significantly outperforming commercial Pt/C in both activity and durability. More broadly, these findings provide a mechanistic basis for the rational design of heterogeneous catalysts and show that structural instability can be converted from a longstanding challenge into a productive route to new catalytic materials.

## Materials and Methods

***Operando* TEM/STEM Investigations** *Operando* STEM/TEM observations were performed on a JEOL GrandARM 300F aberration-corrected microscope. Pt NPs (50 nm, 99.9%, Adamas-Beta, CN), Pd NPs (50 nm, 99.95%, Adamas-Beta, CN) and grinded Rh black (99.9%, Alfa-Aesar, US) were dispersed in ethanol, deposited onto MEMS E-chips (Protochips Inc), and loaded into a gas cell holder, enabling simultaneous heating with flowing gas mixtures. Samples were pretreated under 10% $H_2$/90% Ar ($10^5$ Pa) at 873 K for 30 min[35]. After the pretreatment, a series of ethylene and propane oxidation reactions were conducted in argon balanced flowing mixture (~0.1 sccm) at a total pressure of ~13000 Pa from RT to 873 K. Specific gas compositions are provided in the main text. Exhaust gases were analyzed by MS. Relative changes in associated gases during catalytic reactions were measured by partial pressures at specific atomic mass.

***Ex-situ* TEM/STEM Investigations** Pre- and post-reaction samples were characterized using identical instrumentation. Depth-sensitive HAADF-STEM imaging (**Figure 1**g) employed a 40 mrad semi-convergence angle, equivalent to a 2.17 nm depth of focus[67]. Through-focus series at 2.0 nm increments were captured for analyzing three dimensional $Pt^1$ distribution within GL.

**Near Ambient Pressure (AP) XPS** The *in-situ* NAP-XPS experiments were performed at BL02B01 of the Shanghai Synchrotron Radiation Facility[68]. The bending magnet beamline delivers soft X-rays with a photon flux around $10^{11}$ photons/s at $E/\Delta E = 3700$ and a tightly focused beam spot size (~200 μm × 75 μm) at the sample. Prior to gas introduction, the stainless-steel gas delivery tubes were baked under vacuum and repeatedly flushed with high-purity $O_2$ (99.999%) and $C_2H_4$ (99.999%). The sample temperature, controlled via an infrared laser heating unit, was varied from 300 K to 873 K. Reaction products within the analysis chamber were monitored using a Pfeiffer QMG 250 F1 mass spectrometer connected to the first differential pumping stage. All XPS spectra were calibrated using the Au 4f peak at 84.0 eV binding energy.

**Environmental SEM Investigations** Morphological evolution was monitored using an FEI Helios environmental SEM operated at 15 kV in secondary electron mode. During observation, samples were heated to 873K on a thermal stage with gas flow composition shifting from pure $C_2H_4$ (100 Pa) to $C_2H_4$ (25 Pa) and $O_2$ (75 Pa) at constant total pressure.

**Electrochemical Measurements** All electrochemical measurements were performed using a CHI 760e electrochemical workstation with a standard three-electrode configuration. Gr was purchased from Nanjing XFNANO Materials Tech Co., Ltd, Pt/C (20%Pt) was purchased from Premetek Co., Ltd, and $X^1$/Gr-$X^N$ served as the working electrodes. A Pt wire (1 × 1 cm) and a Hg/HgO electrode were used as the counter and reference electrodes, respectively. The electrolyte was 1 M KOH. A Nafion 117 membrane separated the anode and cathode compartments. All potentials were converted to the reversible hydrogen electrode (RHE) scale using the Nernst equation: $E_{RHE} = E_{Hg/HgO} + 0.0591 \times pH + 0.097$ V. Linear sweep voltammetry (LSV) was conducted at a scan rate of 10 mV $s^{-1}$. All electrochemical data were corrected for solution resistance (*iR* compensation). The solution resistance ($R_s = 1.4\ \Omega$), determined automatically using the CHI760E's "*iR* Compensation" function, was applied to all measurements. CV scans were performed in the non-Faradaic potential window (0.85 to 1.00 V *vs.* RHE) at scan rates of 10, 20, 50, and 100 mV $s^{-1}$ to obtain $C_{dl}$. EIS was measured from $10^5$ Hz to $10^{-2}$ Hz with a 5 mV AC perturbation amplitude. Experimental data were fitted using Zview software to equivalent circuit models R. The economic analysis model is optimized based on our previous work[69].

**Raman Analysis:** Molecular vibrations were excited using 532 nm wavelength light. The exposure time was 20 s. Measurements were performed at different sample locations to ensure reproducibility.

## Author Contributions

Z.-J.W. and Z.C. conceived this project and supervised the research. *In-situ* SEM experiments and analysis were conducted by Z.-J.W.. AP-XPS experiments and analysis were performed by J.C. and Z.L.. The *in-situ* TEM, high-angle annular dark-field STEM and EELS measurements were performed by Z.C., Y.H., Z.Y. and Z.-J.W.. The theoretical simulations were carried out by J.L., L.L. and B.Y.. The electrochemical measurements were performed by Z.F., W.Z. and Zupeng C.. The setup and modification of the experimental apparatus were carried out by Z.-J.W.. Z. C., Y. H., Z. Y. and J. L. contributed equally to this work. Important contributions to the interpretation of the results, conception and writing of the paper were made by Z.C. and Z.-J.W.. All authors participated in the scientific discussion.